# Equivalent Model of Transient Gyrotron Cathode Response

Weiye Xu*, Handong Xu*, Fukun Liu, Xiaojie Wang, Yong Yang, Jian Zhan

*Institute of Plasma Physics, Chinese Academy of Sciences, Hefei, Anhui, China*

*Abstract*—**The gyrotron is the high power millimeter wave source in the electron cyclotron resonance heating system for the tokamak. The cathode power supply is one of the most important ancillary devices for gyrotron. Some interesting transient phenomena about the cathode voltage and the cathode current was found in the gyrotron operation in the electron cyclotron resonance heating system for the Experimental Advanced Superconducting Tokamak. The cathode voltage drops to about 10% of the original value by about 90 μs in the overcurrent case, which is much longer than the 25 μs in the normal case. In order to explain these phenomena, an equivalent circuit model of the magnetron injection gun was proposed. The equivalent circuit is composed of parallel resistors and capacitors, and it can explain the test results very well. Using the equivalent circuit model to analyze gyrotrons may become an effective means of developing and operating an electron cyclotron heating system.**

## I. INTRODUCTION

The electron cyclotron (EC) system is an important plasma heating and current drive method. It is widely used for various fusion experimental devices [1-11], and it is also an important auxiliary heating system for ITER [12, 13].

A 140GHz/4MW electron cyclotron resonance heating (ECRH) system consisting of four gyrotrons for Experimental Advanced Superconducting Tokamak (EAST) is being built in the Institute of Plasma Physics Chinese Academy of Sciences (ASIPP) [14]. Up to now, the first three gyrotrons have been tested, and the fourth gyrotron is being manufactured in CPI. The robust control and protection system, data acquisition system [15], and power measurement system [16] were built for the gyrotrons. In the commissioning process of the first three gyrotrons, many researches related to the gyrotron were done. In recent experiment, the #1 gyrotron oscillation of 980kW/1s, 903kW/10s, 834kW/95s and 650kW/754s were demonstrated, the #2 gyrotron oscillation of 721kW/0.5s, 647kW/2s, 499kW/80s and 406kW/98s were demonstrated, and the #3 gyrotron oscillation of 787kW/20s, 637kW/100s and 559kW/1000s were demonstrated. The output power was measured using calorimetric method [17, 18].

Currently, many researches about gyrotrons such as beam-wave interaction theory [19, 20], design of the gyrotron [21], transient effects of the output wave [22], and transient millimeter-wave signal analysis [23] have been done in the world. However, there is little research on the transient analysis of the cathode voltage, cathode current, anode voltage, anode current, and so on. We have found some interesting phenomena in the gyrotron operation. An equivalent model of the magnetron injection gun was proposed to explain the test results in this paper.

The details of the transient analysis of the voltages and currents are discussed in the following sections. In section 2, the architecture and schematic of the gyrotron and its ancillary systems and the timing sequence of the gyrotron are given. In section3, the negative high voltage power supply for gyrotrons are introduced. Then, in section 4, the measurement and analysis of the transient response of the cathode current and cathode voltage are given. In section 5, the equivalent model of the gyrotron and its role in analysis of gyrotron operation are discussed. Finally, we give the summary in section 6.

## II. THE ANCILLARY SYSTEMS AND THE OPERATION TIMING SEQUENCE OF THE GYROTRON

The gyrotron can not work without its ancillary systems such as the superconducting magnet and its power supply, the collector power supply, the cathode power supply, the anode power supply, the ion pump power supply, and the filament power supply. The ancillary systems and their connections to the gyrotron are presented in Fig. 1. The current limiting resistors are used to limit the maximum current flowing through the cathode and the anode (50 kΩ for the anode and 20 Ω for the cathode). The DC shunts, which are actually resistors with small resistance value, are used to measure anode current and beam current (500 mΩ for the anode current and 1 mΩ for the cathode current). Because the filament is floating on the cathode high voltage, an isolation transformer is used between the filament and its power supply to protect the filament power source.

———————————
∗ Corresponding authors.
E-mail addresses: xuweiye@ipp.cas.cn, xuweiyer@yeah.net (W. Xu), xhd@ipp.cas.cn (H. Xu).

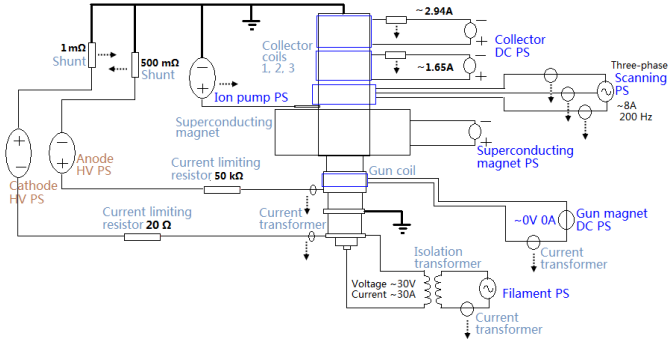

Fig. 1. The gyrotron and its ancillary systems. The 'PS' in this figure is a shorthand for 'power supply'; the 'HV' is a shorthand for 'high voltage'.

The gyrotrons must be operated at a right timing sequence, otherwise the gyrotrons may be damaged. The right timing sequence is shown in Fig. 2. The PLC_Ready signal is an interlock signal from PLC (Programmable Logic Controller). The TriggerIn_-60 signal is sent from EAST center controller, if the rising edge is detected by the timing controller, the timing controller will change the NegHVPre_-60 signal to the high level. In addition, the NegHVPre_-60 signal is sent to the cathode power supply, if the NegHVPre_-60 signal is in the high level, the switchgear of the cathode power supply will be closed. If the switchgear of the cathode power supply is closed, the cathode power supply will send a NegHv_Ready signal to the timing controller. Sixty seconds after the rising edge of the TriggerIn_-60 signal is detected, a rising edge of the TriggerIn_0 signal will be sent to the timing controller. Then the timing controller will turn the NegHv_OnOff signal to be high level immediately (in several nanosecond). Then the IGBTs of the cathode power supply will be closed according to the set time. If the output voltage of the cathode power supply is bigger than 30 kV, the NegHV_OutputState signal will be the high level. Fifty milliseconds after the NegHv_OnOff signal changed to be high level, the timing controller will turn the PosHV_OnOff signal to be high level to turn on the anode power supply. One millisecond after the anode power supply is turned on, the timing controller will detect the Wave_OutputState signal (RF signal) to realize part of the RF protection[16]. If the Wave_OutputState signal goes low, the shutdown procedure will be started, i.e., the NegHv_OnOff signal and the PosHV_OnOff signal will be set to be low level successively (an interval of 2 ms) to shutdown the power supplies to protect the gyrotron. In the process of the gyrotron discharge, if the Ip_D signal (plasma current signal sent from EAST center controller) goes low, the shutdown procedure will be started to protect the EAST tokamak from being damaged by the millimeter wave outputted from the gyrotrons. For safety, anyone of these signals such as PLC_Ready, NegHv_Ready, NegHV_OutputState, Wave_OutputState, and Ip_D goes low, the shutdown procedure will be started.

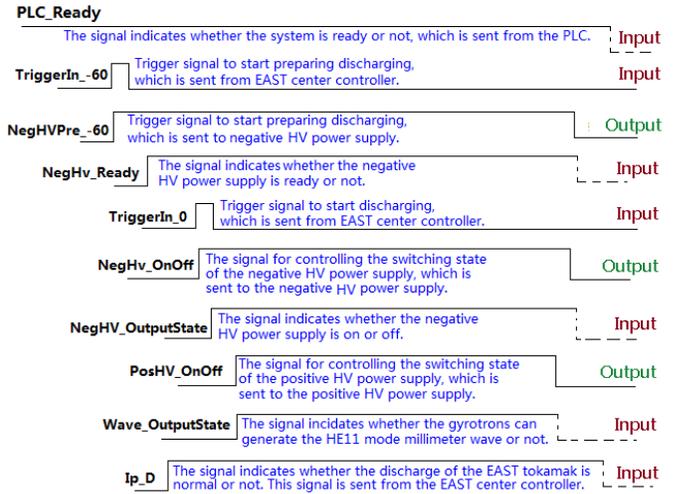

Fig. 2. The operation sequence of the gyrotron. The signals that are marked as 'Input' indicate that the signals are input to the timing controller. The signals that are marked as 'Output' indicate that the signals are output from the timing controller.

## III. Negative High Voltage Power Supply for Gyrotrons

We have developed two cathode high voltage power supplies for four gyrotrons. The cathode high voltage power supplies are using PSM (Pulse Step Modulation) technology [24], which overcomes the shortcomings of the traditional high voltage power supply, such as large single volume, low efficiency, net side low harmonics pollution, lower power factor, larger output ripple, and slower dynamic response. The power supply topology is shown in Fig. 3. In order to protect the gyrotrons, it is necessary to insure that the stored energy of the power supply system is small enough. The energy is mainly stored in the output filter and stray capacitances for PSM modules. The stored energy is very small (<10 J) in our power supply system. In order to verify the protection effect, we have taken a short circuit test. A fuse (whose fusing energy is 10 J) which is connected in series in the loop is still good when the load is shorted, which can prove that the stored output energy of the power supply is less than 10 J. Therefore, we can use this power supply for gyrotron without crowbar.

If the high voltage source receives a turn-off signal or protection signal from the gyrotron control system, the IGBTs will be shut down within several microseconds, thus the connection between the power supply and the gyrotron will be cut off. A test was made to verify the shutdown time of the cathode power source was in several microseconds. A dummy load whose resistance value is 304 Ω was connected to the power source. The power supply shut down when a turn-off signal was send to the power supply. The waveforms of the voltage and the current of the cathode power supply when the IGBTs shut off is shown in Fig. 4. As we can see, the shutdown time of the cathode power source is about 5 µs. It is short enough to protect the gyrotrons.

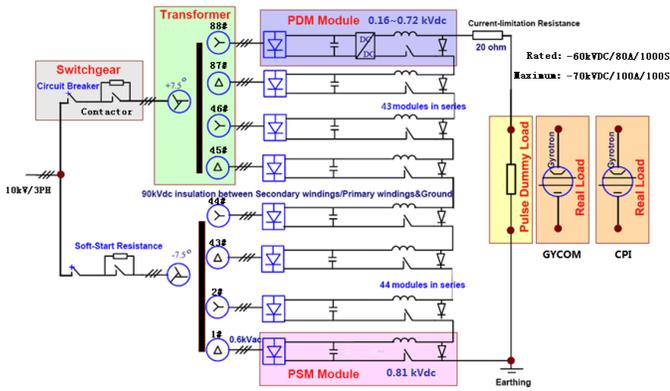

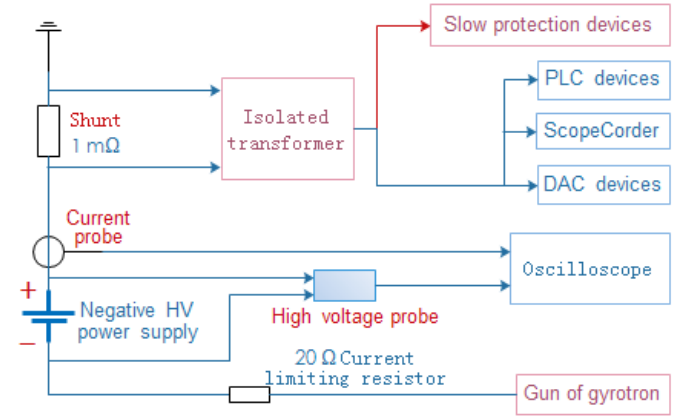

Fig. 3. The schematic of the cathode power supply whose nominal parameter is -60kV/80A/1000s for gyrotrons. The power supply is composed of 87 PSM modules and 1 PDM (Pulse Density Modulation) module.

Fig. 5. The measurement block diagram of the cathode voltage and the cathode current.

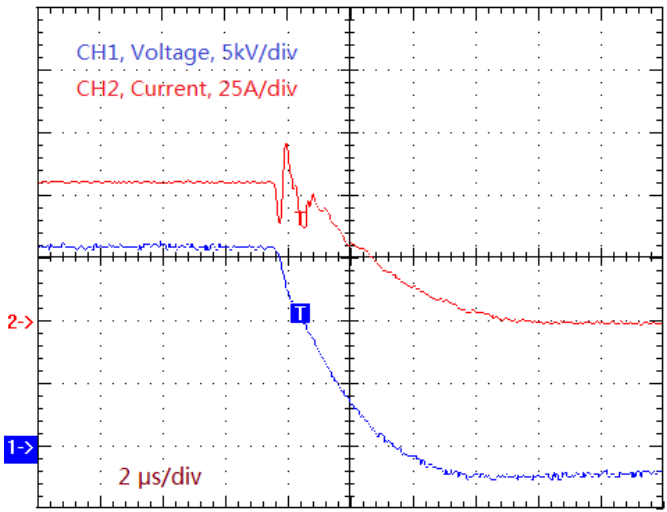

Fig. 4. The waveforms of the voltage and the current when the IGBTs shut off with a dummy load whose resistance value is 304 Ω.

## IV. Measurement and Analysis of the Transient Response of the Cathode Current and the Cathode Voltage

The measurement block diagram of the cathode voltage and the cathode current is shown in Fig. 5. A high voltage probe (Max frequency 20MHz) that is in parallel with the cathode high voltage power supply is used to measure the cathode voltage. A rapid response current probe (Max frequency 400MHz) is used to measure the cathode current. The high voltage probe and the current probe are both connected to the same oscilloscope, which is mainly used to measure the transient response. In addition, a shunt whose resistance is 1 mΩ is used to measure the cathode current. It is used to realize the slow overcurrent protection and to measure the steady-state cathode current.

Fig. 6 and Fig. 7 show the cathode voltage and the cathode current displayed on the oscilloscope when the gyrotron is shutting down. As we can see, the response time of the cathode voltage when the gyrotron shuts down normally is different with the response time of the cathode voltage when the overcurrent protection happens. In the case of normal shutdown, the cathode voltage drops to about 10% of the original value for about 25 μs. In the case of overcurrent, the cathode voltage drops to about 10% of the original value for about 90 μs. Actually, in both two cases, the circuit connection and the measurement method have not changed at all, the only factor of change is the gyrotron. The cathode voltage power supply always has the same shutdown process, the cathode power supply shuts down with the same operation within 6 μs. Therefore, we can infer that some gyrotron parameters changes when an overcurrent happens. The change of the gyrotron parameter causes the change of the drop time of the cathode voltage. We proposed an equivalent model of the gyrotron gun to analyze it.

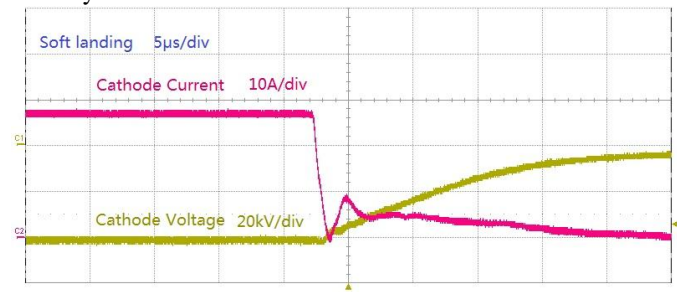

Fig. 6. The cathode voltage and the cathode current when the gyrotron is shutting down normally.

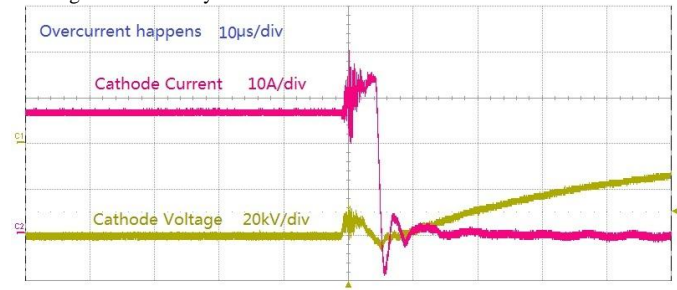

Fig. 7. The cathode voltage and the cathode current when the gyrotron is shutting down with overcurrent.

The equivalent circuit of the gyrotron gun and the auxiliary power supplies can be represented by Fig. 8. Where $R_{rb}$ is the cathode current limiting resistor with the resistance of 20Ω; $R_{ra}$ is the anode current limiting resistor with the resistance of 50kΩ; $R_{sb}$ is the shunt for measuring beam current with the resistance of 1 mΩ; $R_{sa}$ is the shunt for measuring the anode current with the resistance of 500 mΩ; $R_b$ is the equivalent resistor of the electron gun between the cathode and the ground (body); $C_b$ is the equivalent capacitor between the cathode and the ground (body); $R_a$ is the equivalent resistor between the anode and the cathode; $C_a$ is the equivalent capacitor between the anode and the cathode.

The resistance of $R_b$ is related to the voltage across $C_b$ and the voltage across $C_a$ (the sum of the absolute value of the cathode voltage and the absolute value of the anode voltage) and the power of the filament power supply. In the case of the normal shutdown process of the cathode voltage, the anode voltage has been reduced to 0, so the value of $R_b$ is just related to the voltage across $C_b$ (the cathode voltage) and the power of the filament power supply. Table 1 shows the examples.

TABLE I
The Resistance of Rb Along with the Cathode Voltage, the Anode Voltage, and the Filament Power.

| Cathode voltage [kV] | Anode voltage [kV] | Filament power [W] | Cathode Current [A] | $R_b$ [kΩ] |
|---|---|---|---|---|
| -45 | 0 | 1135.32 | 32.2 | 1.3975 |
| -44 | 0 | 1126.09 | 30.6 | 1.4379 |
| -44 | 0 | 1113.26 | 30.1 | 1.4618 |
| -43 | 0 | 1111.77 | 29.2 | 1.4726 |
| -42 | 0 | 1099.59 | 26.5 | 1.5849 |
| -42 | 0 | 1097.65 | 25.9 | 1.6216 |
| -40 | 0 | 1097.65 | 25.6 | 1.5625 |
| -35 | 0 | 1097.65 | 24.5 | 1.4286 |
| -20 | 0 | 1097.65 | 20.0 | 1.0000 |
| -10 | 0 | 1097.65 | 12.0 | 0.8333 |
| -42 | 5 | 1099.59 | 27.3 | 1.5385 |
| -42 | 16 | 1099.59 | 29.0 | 1.4483 |
| -42 | 19 | 1099.59 | 29.4 | 1.4286 |

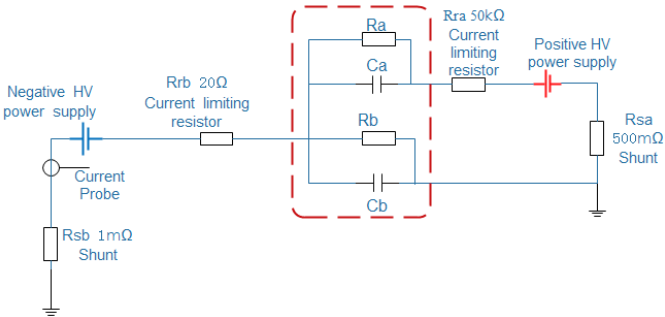

Fig. 8. The equivalent model of the gyrotron gun and the auxiliary power supplies. The equivalent model of the gun is in the red dashed box.

We analyzed the relationship between the value of $R_b$ and the cathode voltage $u_b$ in the situation where the anode voltage is zero and the filament power is 1097.65 W. The relationship between the value of $R_b$ and the absolute value of the cathode voltage $u_b$ are shown in Fig. 9. The exponential fitting function is,

$$R_b = -359.6 + 1005.5 \cdot \exp(1.6 \times 10^{-5} u_b). \quad (1)$$

Where the unit of $u_b$ is Volt, and the unit of $R_b$ is Ohm.

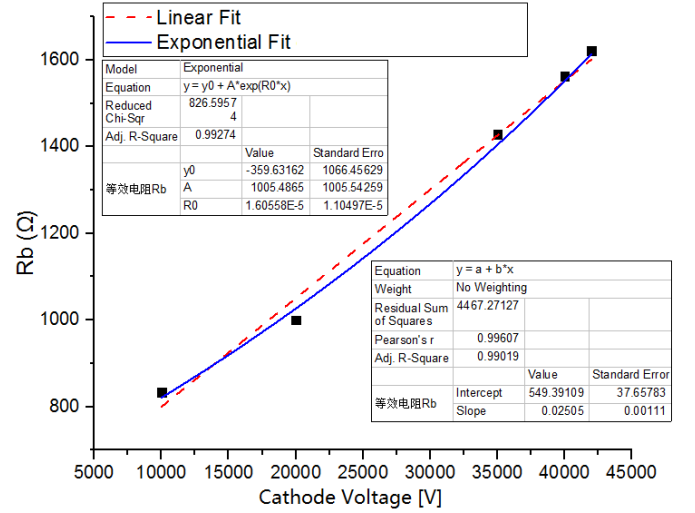

Fig. 9. The value of $R_b$ along with the cathode voltage.

When the cathode power supply shuts down, a capacitor discharge circuit is formed by $C_b$ and $R_b$. That is a zero-input response. Assume the initial voltage across the capacitor $C_b$ is $U_b$ (a positive value in the unit of Volt), the Laplace transform equivalent circuit is shown in Fig. 10. The voltage across the capacitor $C_b$ is,

$$u(s) = \frac{U_b}{s} - \frac{\frac{U_b}{s} \cdot \frac{1}{sC_b}}{\frac{1}{sC_b} + R_b} = \frac{U_b}{s} - \frac{U_b}{s(1 + sC_b R_b)} = \frac{U_b}{\frac{1}{C_b R_b} + s}. \quad (2)$$

Using inverse Laplace transform, we can get,

$$u(t) = U_b \cdot \exp(-\frac{1}{C_b R_b} t). \quad (3)$$

Where $u(t) = u_b$. By solving equation (1) and equation (3), assume $u(t) = u_b = u_t$, we can get,

$$u_t = U_b \cdot \exp(-\frac{1}{C_b(-359.6 + 1005.5 \cdot \exp(1.6 \times 10^{-5} u_t))} t). \quad (4)$$

Take natural logarithm at both ends of the above equation, and simplify the equation,

$$359.6(\ln u_t - \ln U_b) + 1005.5(\ln U_b - \ln u_t) \exp(1.6 \times 10^{-5} u_t) = \frac{t}{C_b}. \quad (5)$$

Let $U_b = 41000$ V, $u_t = 4100$ V, we can get the time used for the cathode voltage decreasing to about 10% of the original value,

$$t \approx 1644.2 C_b. \tag{6}$$

As shown in Fig. 6, in the case of normal shutdown, the cathode voltage drops to about 10% of the original value for about 25 μs. So, the equivalent capacitance is,

$$C_b \approx \frac{t}{1644.2} \approx 15.2 \text{ nF}. \tag{7}$$

For the current flowing through $R_b$, we assume the direction of the current shown in Fig. 10 is positive, then,

$$i(t) = -C_b \frac{du_t}{dt} = \frac{U_b}{R_b} \cdot \exp(-\frac{1}{C_b R_b} t) \tag{8}$$

Since $R_b$ varies with the voltage $u_t$, the equation (3) is brought into the above equation to obtain the relationship between the current $i(t)$ and the voltage $u_t$,

$$i(t) = \frac{u_t}{R_b} = \frac{u_t}{-359.6 + 1005.5 \cdot \exp(1.6 \times 10^{-5} u_t)}. \tag{9}$$

The value of $i(t)$ is always positive, indicate that the real direction of the current is same as the direction shown in Fig. 10, i.e., the real direction of the current is same as the direction at time 0.. The relationship between $i(t)$ and $u_t$ is shown in Fig. 11. When $u_t$=41000 V, $i(t)$≈26 A=i(0-); when $u_t$=30000, $i(t)$≈24 A; when $u_t$=20000 V, $i(t)$≈20 A; when $u_t$=0 V, $i(t)$≈0 A. The value of $i(t)$ decreases with the decrease of $u_t$, and the drop time is almost the same. It should be noted that the current $i(t)$ is the current flowing through the resistor $R_b$, it is not the same one measured by a current probe shown in Fig. 5, and it is hardly to be measured.

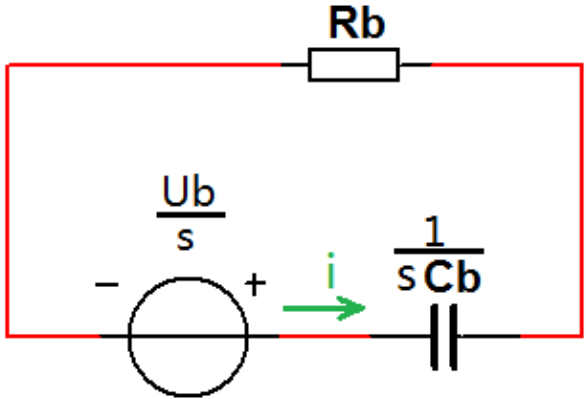

Fig. 10. The Laplace transform equivalent circuit when the cathode power supply shuts down.

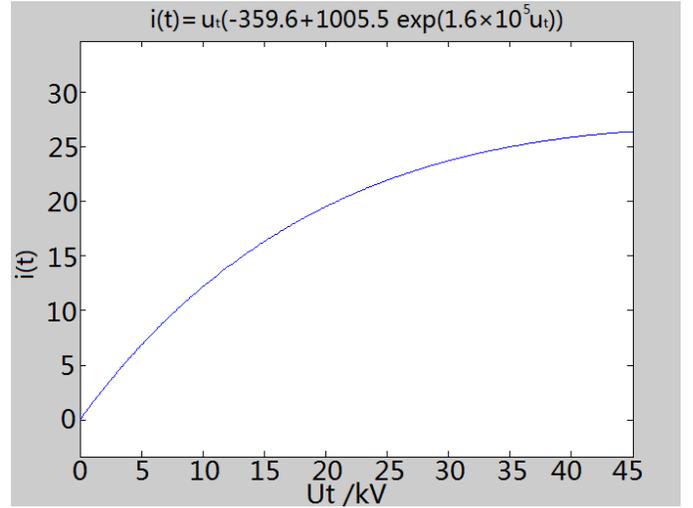

Fig. 11. The current flowing through $R_b$ as a function of the voltage $u_t$.

The cathode current will increase suddenly (overcurrent happens) when a small discharge occurs or some other factors change inside the gyrotron gun. It is dangerous for gyrotron. Therefore, if an overcurrent happens, the protection system will shut down the anode power supply and the cathode power supply at the same time. As we can see in Fig. 7, about 5 μs after overcurrent happens, the voltage began to turn off. Therefore, after overcurrent happens, the energy transmitted to the gyrotron is,

$$W \approx \int pdt \approx \int_0^{5\mu} uidt \approx 41\text{kV} \times 35\text{A} \times 5\mu\text{s} \approx 7.2 \text{ J}. \tag{10}$$

In the situation where overcurrent happens, the anode voltage starts to decrease at the same time as the cathode voltage. That is, the anode voltage is not zero when the cathode power supply shuts down. But we can see from Table 1, the value of $R_b$ is less affected by the anode voltage. For the sake of simplicity, we ignore the anode voltage in the overcurrent situation. As we can see in Fig. 7, the cathode voltage drops to about 10% of the original value for about 90 μs, which is much longer than the time in the normal situation. The equivalent capacitance is,

$$C_b \approx \frac{t}{1644.2} \approx 54.7 \text{ nF}. \tag{11}$$

The equivalent capacitance $C_b$ increases may be due to the discharge between the cathode and ground, which leads to the increase of the drop time of the cathode voltage.

In order to further verify the above assumptions, we try to add a crowbar short-circuit switch at both ends of the cathode voltage source. When the cathode voltage source is turned off, the crowbar short-circuit switch is automatically closed and the equivalent circuit is shown in Fig. 12. When the crowbar switch is closed, the capacitor $C_b$ will be discharged through the $R_{rb}$. The voltage across the capacitor $C_b$ is,

$$u(t) = U_b \cdot \exp(-\frac{1}{C_b R_{rb}} t). \tag{12}$$

The cathode voltage drops to about 10% of the original value for about,

$$2.3R_{rb}C_b \approx 0.7 \text{ μs.} \tag{13}$$

The actual cathode voltage waveform is shown in Fig. 13, and it can be seen that the voltage falling edge is indeed about 0.7 μs. However, because the voltage source turns off too fast, the oscillation of the voltage will be generated [25]. The oscillation occurs due to the signal transmission and reflection between the left end of the $C_b$ and the ground.

The current flowing through $R_{rb}$ is,

$$i(t) = -C_b \frac{du_t}{dt} = \frac{U_b}{R_{rb}} \cdot \exp\left(-\frac{1}{C_b R_{rb}}t\right). \tag{14}$$

When $U_b$=41 kV, the current is about 2.05 kA at time zero, and the current direction flowing through $R_{rb}$ is opposite to the initial direction. Then it decreases to about 205 A after 0.7 μs, decreases to about 20.5 A after 1.4 μs. It can be seen from the beam current signal in Fig. 13 that the current signal does rush to a large value, but since the oscilloscope's preset signal amplitude range is small, we do not see how much the specific maximum value is. It can be seen from Fig. 13 that the current signal drops to near 0 A for about 1 μs, which is in line with the theoretical expectation. However, owing to the current edge is too fast, the current signal is transmitted back and forth on the signal line, the current oscillation occurs.

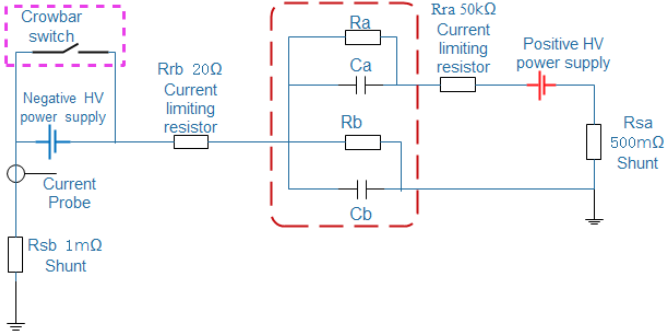

Fig. 12. The equivalent model of the gyrotron gun and its power supplies with the crowbar switch.

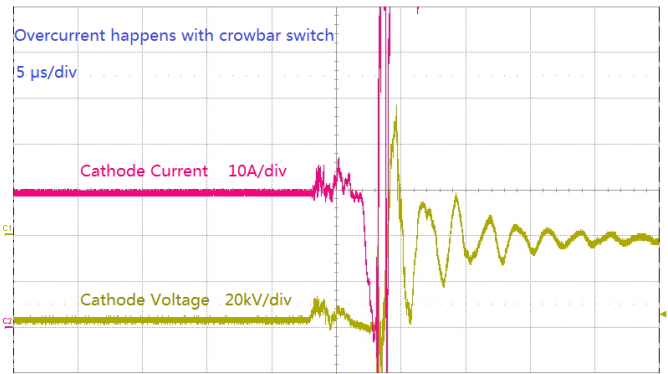

Fig. 13. The cathode voltage and the cathode current in the case that the gyrotron is shutting down when overcurrent happens with crowbar switch.

The transient response of the anode current and anode voltage when the power supply shuts down is similar to that of the cathode. The equivalent resistor $R_a$ shown in Fig. 8 is related to the voltage across $C_a$ (the sum of the absolute value of the cathode voltage and the absolute value of the anode voltage). The resistance of $R_a$ decreases as the anode voltage increases. $R_a$ is probably several megohms; $C_a$ is probably several picofarads, and $C_a$ may increase in the overcurrent condition. The time of the anode voltage drops to about 10% of the original value in the overcurrent case may be longer than that in the normal case. If just the cathode voltage is applied to the gyrotron, and the anode is not connected to the anode power supply or ground, the potential on the anode will be equal to the potential on the cathode. If the anode potential is not equal to the cathode potential, there may be an equivalent resistor between the anode and the ground. More detailed analysis and test will be made in the future.

## V. The Equivalent Circuit Model of the Gyrotron and its Role in Analysis of Gyrotron Operation

We proposed an equivalent circuit model of the gyrotron gun in part 4. Actually, the gyrotrons can be analyzed using the same equivalent circuit model, which is shown in Fig. 14. Where, $R_b$ is the equivalent resistor of the gyrotron between the cathode and the ground; $C_b$ is the equivalent capacitor between the cathode and the ground; $R_a$ is the equivalent resistor between the anode and the cathode; $C_a$ is the equivalent capacitor between the anode and the cathode.

In the normal discharge of gyrotron, the value of $R_b$ is related with the filament power, the cathode voltage, the anode voltage, the pulse duration, etc. Usually, the beam current gradually decreases as the pulse time becomes longer [26], that is, the value of $R_b$ increases as the pulse time becomes longer. In the security area, the value of $R_b$ decreases as the filament power becomes higher, as the anode voltage becomes higher, or as the cathode voltage becomes lower. The value of $R_a$ is related with the magnetic field strength in cathode region. When the magnetic field strength in the cathode region decreases, the anode current increases, resulting in a decrease in $R_a$. For simplicity, the value $C_a$ and $C_b$ may be seen as a constant.

The characteristics of the anode voltage, anode current, cathode voltage, and cathode current when the cathode power supply or the anode power supply is turned off can be analyzed using the method shown in part 4.

The beam current $i_b$ can be calculated by,

$$i_b = \frac{u_b}{R_b}, \tag{15}$$

where, $u_b$ is the cathode voltage.

The anode current $i_a$ can be calculated by,

$$i_a = \frac{u_a + |u_b|}{R_a}, \tag{16}$$

where, $u_a$ is the anode voltage.

The gyrotron output RF power $P_{rf}$ can be calculated by,

$$P_{rf} = \frac{u_b^2}{R_b}\eta, \tag{17}$$

where, $\eta$ is the gyrotron efficiency, it is related with the cathode voltage, the anode voltage, the magnetic field, the filament power, and so on [27]. The gyrotron efficiency can also be measured.

In the future, all the parameters of the equivalent circuit model of the gyrotron may be measured and provided by the gyrotron manufacturer. Therefore, the gyrotron users like the institutes for fusion research can use the gyrotrons more conveniently. That is useful for fusion heating research.

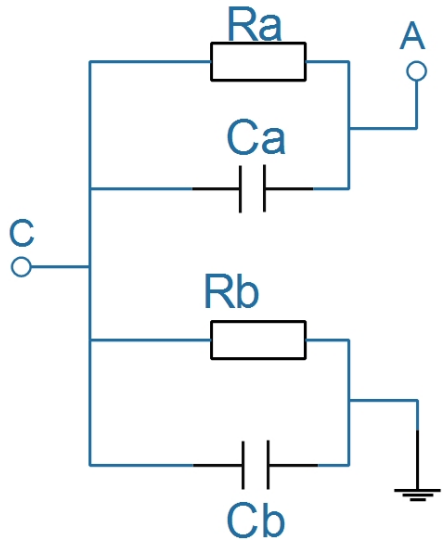

Fig. 14. The simplified equivalent circuit model of the gyrotron. 'C' indicates the cathode power supply terminal, and 'A' indicates the anode power supply terminal.

## VI. CONCLUSION

The gyrotron is the key part of the ECRH system. The gyrotron is a sophisticated vacuum device, and many ancillary systems are needed to assist it in working. The cathode power supply and the anode power supply are the most important ancillary devices. The gyrotrons must be operated at a right timing sequence, otherwise the gyrotrons may be damaged. During the experiments of the gyrotrons, we found that the waveforms of the cathode voltage and the cathode current vary with different conditions. The cathode voltage drops to about 10% of the original value by about 90 μs in the overcurrent case, which is much longer than the 25 μs in the normal case. An equivalent circuit of the gun of the gyrotron is proposed to analyze the transient phenomena about the cathode voltage and the cathode current. The equivalent circuit is composed of parallel resistors and capacitors, and it can explain the test results well. We also simply predicted the response of the anode voltage and current according to the equivalent circuit model. More detailed test will be made in the future. The proposed equivalent circuit model of the gyrotron in this paper is meaningful. We can analyze the gyrotron like a chip using the equivalent circuit model. That may become an effective means of developing and operating an electron cyclotron system in the future.

## ACKNOWLEDGMENTS

This work was supported in part by the National Key R&D Program of China under Grant 2017YFE0300401 and the National Magnetic Confinement Fusion Science Program of China under Grant 2015GB102003 and Grant 2015GB103000.

The authors greatly appreciate the experts from GA, CPI, and GYCOM for the cooperation in the development of the ECRH project on EAST.

## REFERENCES

[1] V. Erckmann, H. Braune, G. Gantenbein, J. Jelonnek, W. Kasparek, H.P. Laqua, C. Lechte, N.B. Marushchenko, G. Michel, B. Plaum, M. Thumm, M. Weissgerber, R. Wolf, W.-X.E. Teams, ECRH and W7-X: An intriguing pair, AIP Conference Proceedings, 1580 (2014) 542-545.

[2] M. Lennholm, G. Giruzzi, A. Parkin, F. Bouquey, H. Braune, A. Bruschi, E. de la Luna, G. Denisov, T. Edlington, D. Farina, J. Farthing, L. Figini, S. Garavaglia, J. Garcia, T. Gerbaud, G. Granucci, M. Henderson, L. Horton, W. Kasparek, P. Khilar, M. Jennison, N. Kirneva, D. Kislov, A. Kuyanov, X. Litaudon, A.G. Litvak, A. Moro, S. Nowak, V. Parail, B. Plaum, F. Rimini, G. Saibene, A. Sips, C. Sozzi, P. Spah, E. Trukhina, A. Vaccaro, V. Vdovin, J.-E.C.S. Ctr, ECRH for JET: A feasibility study, Fusion Eng Des, 86 (2011) 805-809.

[3] J. Stober, A. Bock, H. Hohnle, M. Reich, F. Sommer, W. Treutterer, D. Wagner, L. Gianone, A. Herrmann, F. Leuterer, F. Monaco, M. Marascheck, A. Mlynek, S. Muller, M. Munich, E. Poli, M. Schubert, H. Schutz, H. Zohm, W. Kasparek, U. Stroth, A. Meier, T. Scherer, D. Strauss, A. Vaccaro, J. Flamm, M. Thumm, A. Litvak, G.G. Denisov, A.V. Chirkov, E.M. Tai, L.G. Popov, V.O. Nichiporenko, V.E. Myasnikov, E.A. Soluyanova, S.A. Malygin, A.U. Team, ECRH on ASDEX Upgrade - System Status, Feed-Back Control, Plasma Physics Results -, Epj Web Conf, 32 (2012).

[4] G. Taylor, R.A. Ellis, R.W. Harvey, J.C. Hosea, A.P. Smirnov, ECRH/EBWH System for NSTX-U, Epj Web Conf, 32 (2012) 02014.

[5] A.V. Melnikov, L.G. Eliseev, S.V. Perfilov, V.F. Andreev, S.A. Grashin, K.S. Dyabilin, A.N. Chudnovskiy, M.Y. Isaev, S.E. Lysenko, V.A. Mavrin, M.I. Mikhailov, D.V. Ryzhakov, R.V. Shurygin, V.N. Zenin, T.-. Team, Electric potential dynamics in OH and ECRH plasmas in the T-10 tokamak, Nucl Fusion, 53 (2013).

[6] X.T. Ding, W. Chen, L.M. Yu, S.Y. Chen, J.Q. Dong, X.Q. Ji, Z.B. Shi, Y. Zhou, Y.B. Dong, X.L. Huang, Z.W. Xia, X.Y. Song, X.M. Song, J. Zhou, J. Rao, M. Huang, B.B. Feng, Y. Huang, Y. Liu, L.W. Yan, Q.W. Yang, X.R. Duan, H.-A.E. Team, Energetic Particle Physics Experiments With High Power ECRH on HL-2A, Epj Web Conf, 32 (2012).

[7] G.P. Canal, B.P. Duval, F. Felici, T.P. Goodman, J.P. Graves, A. Pochelon, H. Reimerdes, O. Sauter, D. Testa, T. Team, Fast seeding of NTMs by sawtooth crashes in TCV and their preemption using ECRH, Nucl Fusion, 53 (2013).

[8] K. Hada, K. Nagasaki, K. Masuda, S. Kobayashi, S. Ide, A. Isayama, K. Kajiwara, One-Dimensional Analysis of Ecrh-Assisted Plasma Start-up in Jt-60sa, Fusion Sci Technol, 67 (2015) 693-704.

[9] H. Igami, S. Kubo, T. Shimozuma, Y. Yoshimura, H. Takahashi, S. Kamio, S. Kobayashi, S. Ito, Y. Mizuno, K. Okada, R. Makino, S. Ogasawara, K. Kobayashi, M. Osakabe, K. Nagasaki, H. Idei, T. Mutoh, L.H.D.E. Grp, Recent Upgrading of ECRH System and Studies to Improve ECRH Performance in the LHD, in: S. Kubo (Ed.) Ec18 - 18th Joint Workshop on Electron Cyclotron Emission and Electron Cyclotron Resonance Heating, E D P Sciences, Cedex A, 2015.

[10] M. Cengher, X. Chen, R. Ellis, Y. Gorelov, J. Lohr, C. Moeller, D. Ponce, A. Torrezan, Advances in technology and high power performance of the ECH system on DIII-D, Fusion Eng Des, (2017).

[11] Y.S. Bae, J. Decker, J.H. Jeong, K.D. Lee, Study of Synergetic Effect of X2 and X3 EC Wave in KSTAR, in: EPJ Web of Conferences, 2015.

[12] M. Henderson, G. Saibene, C. Darbos, D. Farina, L. Figini, M. Gagliardi, F. Gandini, T. Gassmann, G. Hanson, A. Loarte, T. Omori, E. Poli, D. Purohit, K. Takahashi, The targeted heating and current drive applications for the ITER electron cyclotron system, Phys Plasmas, 22 (2015) 021808.

[13] T. Omori, F. Albajar, T. Bonicelli, G. Carannante, M. Cavinato, F. Cismondi, C. Darbos, G. Denisov, D. Farina, M. Gagliardi, F. Gandini, T. Gassmann, T. Goodman, G. Hanson, M.A. Henderson, K. Kajiwara, K. McElhaney, R. Nousiainen, Y. Oda, A. Oustinov, D. Parmar, V.L. Popov, D. Purohit, S.L. Rao, D. Rasmussen, D.M.S. Ronden, G. Saibene, K.

Sakamoto, F. Sartori, T. Scherer, N.P. Singh, D. Strauß, K. Takahashi, Progress in the ITER electron cyclotron heating and current drive system design, Fusion Eng Des, 96–97 (2015) 547-552.

[14] H. Xu, X. Wang, F. Liu, J. Zhang, Y. Huang, J. Shan, D. Wu, H. Hu, B. Li, M. Li, Y. Yang, J. Feng, W. Xu, Y. Tang, W. Wei, L. Xu, Y. Liu, H. Zhao, J. Lohr, Y.A. Gorelov, J.P. Anderson, W. Ma, Z. Wu, J. Wang, L. Zhang, F. Guo, H. Sun, X. Yan, T. East, Development and Preliminary Commissioning Results of a Long Pulse 140 GHz ECRH System on EAST Tokamak (Invited), Plasma Science and Technology, 18 (2016) 442-448.

[15] W. Xu, H. Xu, F. Liu, F. Hou, Z. Wu, Data acquisition system for electron cyclotron resonance heating on EAST tokamak, Fusion Eng Des, 113 (2016) 119-125.

[16] W. Xu, H. Xu, F. Liu, Y. Tang, Z. Wu, X. Wang, J. Wang, J. Feng, Millimeter Wave Power Monitoring in EAST ECRH System, IEEE Access, 4 (2016) 5809-5817.

[17] L. Gorelov, J. Lohr, R. Baity Jr, P. Cahalan, R. Callis, D. Ponce, H. Chiu, Gyrotron power balance based on calorimetric measurements in the DIII-D ECH system, in: Fusion Engineering, 2003. 20th IEEE/NPSS Symposium on, IEEE, 2003, pp. 546-548.

[18] W. Xu, H. Xu, F. Liu, J. Wang, X. Wang, Y. Hou, Calorimetric power measurements in the EAST ECRH system, Plasma Science and Technology, 19 (2017) 105602.

[19] S.G. Liu, The kinetic theory of electron cyclotron resonance maser, Scientia Sinica, 22 (1979) 901-911.

[20] E. Borie, B. Jodicke, Comments on the linear theory of the gyrotron, Ieee T Plasma Sci, 16 (1988) 116-121.

[21] M.V. Kartikeyan, E. Borie, O. Drumm, S. Illy, B. Piosczyk, M. Thumm, Design of a 42-GHz 200-kW gyrotron operating at the second harmonic, Microwave Theory & Techniques IEEE Transactions on, 52 (2004) 686-692.

[22] A.T. Lin, Transient Effects In High Current Gyrotrons, in, 1988, pp. 8-17.

[23] A. Schlaich, G. Gantenbein, J. Jelonnek, M. Thumm, Transient Millimeter-Wave Signal Analysis With Unambiguous RF Spectrum Reconstruction, Ieee T Microw Theory, 61 (2013) 4660-4666.

[24] Z. Yang, J. Zhang, Y. Huang, X. Hao, Q. Zhao, F. Guo, The Analysis and Design of High Voltage DC Power Supply of ECRH for EAST, Nuclear Science and Technology, 01 (2013) 9-17.

[25] H.W. Johnson, M. Graham, High-Speed Digital Design: A Handbook of Black Magic, Prentice-Hall, Englewood Cliffs, NJ, 1993.

[26] T.E. Harris, Active heater control and regulation for the Varian VGT-80 11 gyrotron, in: [Proceedings] The 14th IEEE/NPSS Symposium Fusion Engineering, 1991, pp. 130-131 vol.131.

[27] G.S. Nusinovich, M.K.A. Thumm, M.I. Petelin, The Gyrotron at 50: Historical Overview, J Infrared Millim Te, 35 (2014) 325-381.